\documentclass[11pt]{article}
\usepackage{moriond,epsfig}
\def\Journal#1#2#3#4{{#1} {\bf #2}, #3 (#4)}

\def\NPB{{\em Nucl. Phys.} B}
\def\PLB{{\em Phys. Lett.}  B}

\def\be{\begin{equation}}
\def\ee{\end{equation}}
\def\bea{\begin{eqnarray}}
\def\eea{\end{eqnarray}}

\begin{document}
\vspace*{4cm}
\title{ENHANCED DIAGRAMS IN HIGH ENERGY HADRONIC AND NUCLEAR SCATTERING}

\author{ S.S. OSTAPCHENKO}

\address{Forschungszentrum Karlsruhe, Institut f\"ur Kernphysik, 
76021 Karlsruhe, Germany  \\
         Moscow State University, 
	 D.V. Skobeltsyn Institute of Nuclear Physics, 
119992 Moscow, Russia}

\maketitle\abstracts{
High energy hadronic and nuclear interactions are described within
Gribov's Reggeon scheme. An approach to re-summation of enhanced
Pomeron graphs is proposed. The latter is applied to develop a new
Monte Carlo model which treats non-linear interaction 
effects explicitely in individual hadronic collisions. 
On the other hand, we discuss a possible generalization of the scheme,
which allows to account also for pQCD effects. In particular, this offers
a possibility to fix model parameters on the  basis of data
on hadron-hadron cross sections, hadronic multi-particle production, 
and on total and diffractive proton structure functions.}

\section{Introduction}

Investigations of high energy hadronic and nuclear interactions remains an 
interesting and grateful field of research. An important ingredient of such
studies is the development of corresponding Monte Carlo (MC) models-generators,
the latter being extensively used for projecting new experiments, analyzing
and interpreting measured data.

\begin{figure}[ht]
\begin{center}
\includegraphics[
  width=7cm,
  height=2.5cm]{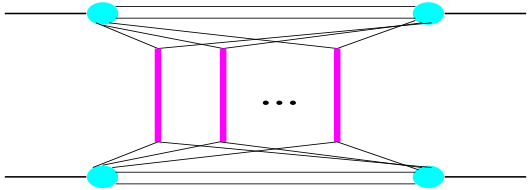}
  \end{center}
\caption{A general multi-Pomeron contribution to hadron-hadron scattering
amplitude. Elementary scattering processes (vertical thick lines)
are described as Pomeron exchanges.\label{multiple}}
\end{figure}

Still, despite a significant progress in QCD over the past few decades, only 
a phenomenological treatment is generally possible for minimum-bias
hadron-hadron (hadron-nucleus, nucleus-nucleus) collisions, typically being
based (explicitely or implicitly) on the  Gribov's Reggeon approach.~\cite{gri68}
In the latter scheme a high energy hadron-hadron collision is described as a 
multiple scattering process, where elementary re-scatterings, corresponding
to microscopic parton cascades, are treated 
phenomenologically as Pomeron exchanges -- Fig.~\ref{multiple}.%

The Pomeron exchange amplitude is typically chosen as~\cite{kai82}
\begin{eqnarray}
f_{ad}^{{\rm P}}(s,b)&=&
\frac{i\,\gamma_{a}\,\gamma_{d}\,(s/s_{0})^{\alpha_{{\rm P}}(0)-1}}
{\lambda_{ad}^{{\rm P}}(s)}
\: \exp \!\left(-\frac{b^{2}}{4\,\lambda_{ad}^{{\rm P}}(s)}\right) \label{f-pom} \\
\lambda_{ad}^{{\rm P}}(s)&=& R_{a}^{2}+R_{d}^{2}+\alpha_{{\rm P}}'(0)\,
 \ln (s/s_{0}),
 \end{eqnarray}
where  $s_{0}\simeq 1$ GeV$^{2}$ is the hadronic mass scale, $\alpha_{{\rm P}}(0)$
and $\alpha_{{\rm P}}'(0)$ are the intercept and the slope of the
Pomeron Regge trajectory, and $\gamma_{a}$, $R_{a}^{2}$ are the
coupling and the slope of Pomeron-hadron $a$ interaction vertex.
Assuming an over-critical Pomeron trajectory ($\alpha_{{\rm P}}(0)>1$) it is
characterized by a power-like energy increase and a Gaussian impact parameter shape,
with the corresponding width (slope) rising with energy.

Using the optical theorem and calculating various unitarity cuts of elastic scattering 
diagrams of  Fig.~\ref{multiple} according to the so-called Ab\-ram\-ovskii-Gribov-Kancheli
 (AGK) cutting rules,~\cite{agk} one can obtain expressions for total and inelastic
 cross sections as well as for relative probabilities of particular interaction
 configurations, e.g., for a given number of elementary inelastic processes
 ("cut" Pomerons), all being expressed via the Pomeron amplitude $f_{ad}^{{\rm P}}(s,b)$. 
 Furthermore, identifying such "cut" Pomeron processes with string formation and break-up
 allowed to propose powerful model approaches, like Quark-Gluon String or Dual Parton
models,~\cite{kai82} which in turn opened the way to develop corresponding MC generators 
of hadronic and nuclear collisions.
 
Such a scheme is characterized by a great simplicity and flexibility,
a general high energy collision being just a superposition of a number of elementary
processes -- Pomeron exchanges, by small number of adjustable parameters,~\cite{kai82} and by a
 parameter-free generalization to hadron-nucleus and nucleus-nucleus case.~\cite{kal93}
 Nevertheless, its validity is subject to condition that such elementary
 processes proceed independently of each other.

\section{Enhanced Pomeron Diagrams}

The above-mentioned condition is not expected to be valid in
the ``dense'' regime, i.e.~in the limit of high energies $s$
and small impact parameters $b$ of the interaction. There, a large number
of elementary scattering processes occurs and corresponding underlying
parton cascades largely overlap and interact with each other.~\cite{glr} 
Such effects are traditionally described by so-called enhanced 
diagrams, which involve Pomeron-Pomeron interactions.~\cite{kan73,car74}

Recently, a re-summation procedure for higher orders of such diagrams has been
proposed,~\cite{bon01} however, taking into consideration only triple-Pomeron coupling.
Here we shall rather stay close to the $\pi$-meson dominance approach,~\cite{car74}
where all multi-Pomeron vertexes 
have been expressed via the triple-Pomeron coupling constant $r_{3{\rm P}}$
 and where an asymptotic re-summation has been proposed.~\cite{kai86}

\begin{figure}[ht]
\begin{center}
\includegraphics[
  width=3.5cm,
  height=2.5cm]{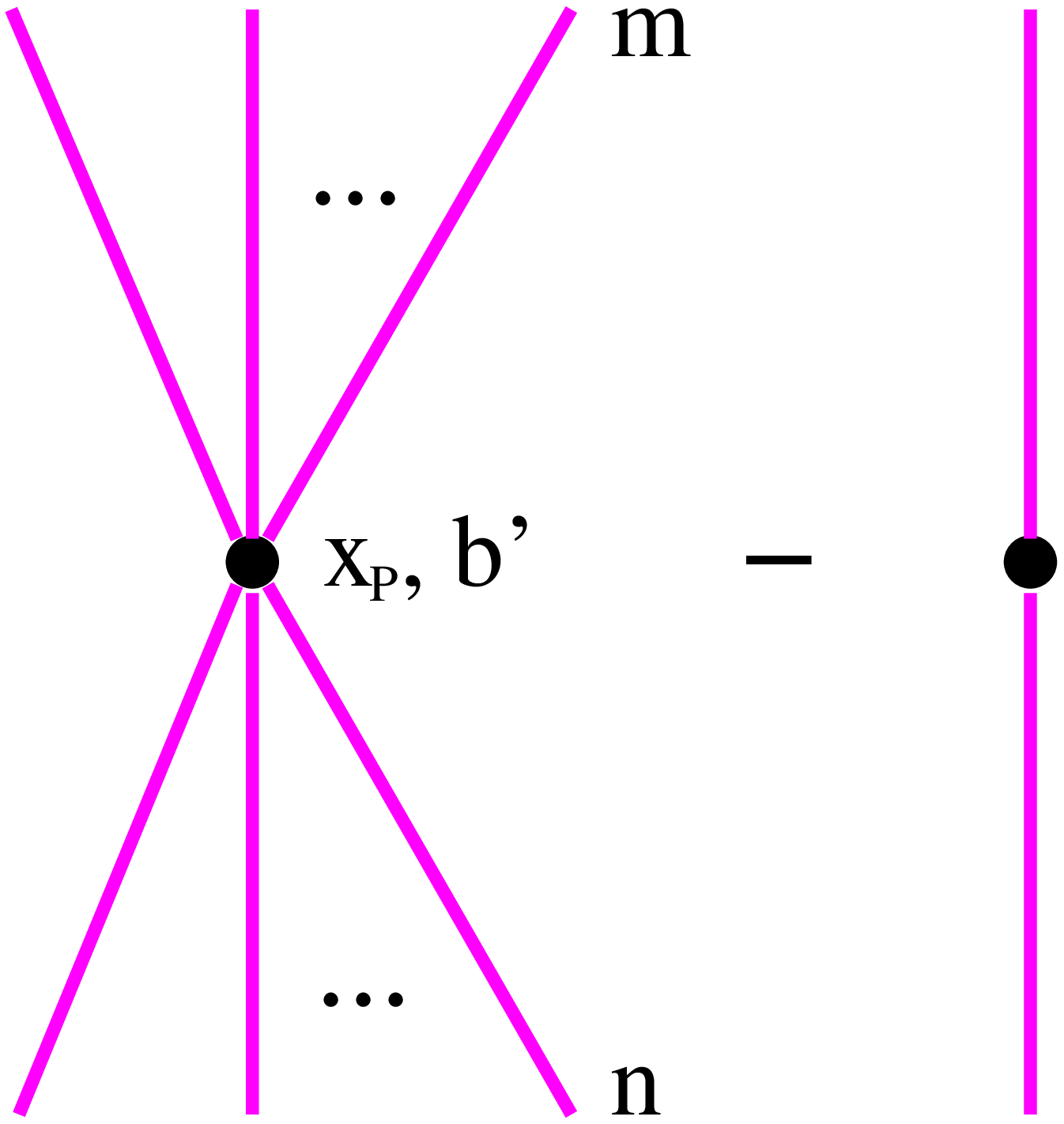}
  \hspace{1cm}
\includegraphics[
  width=5cm,
  height=2.5cm]{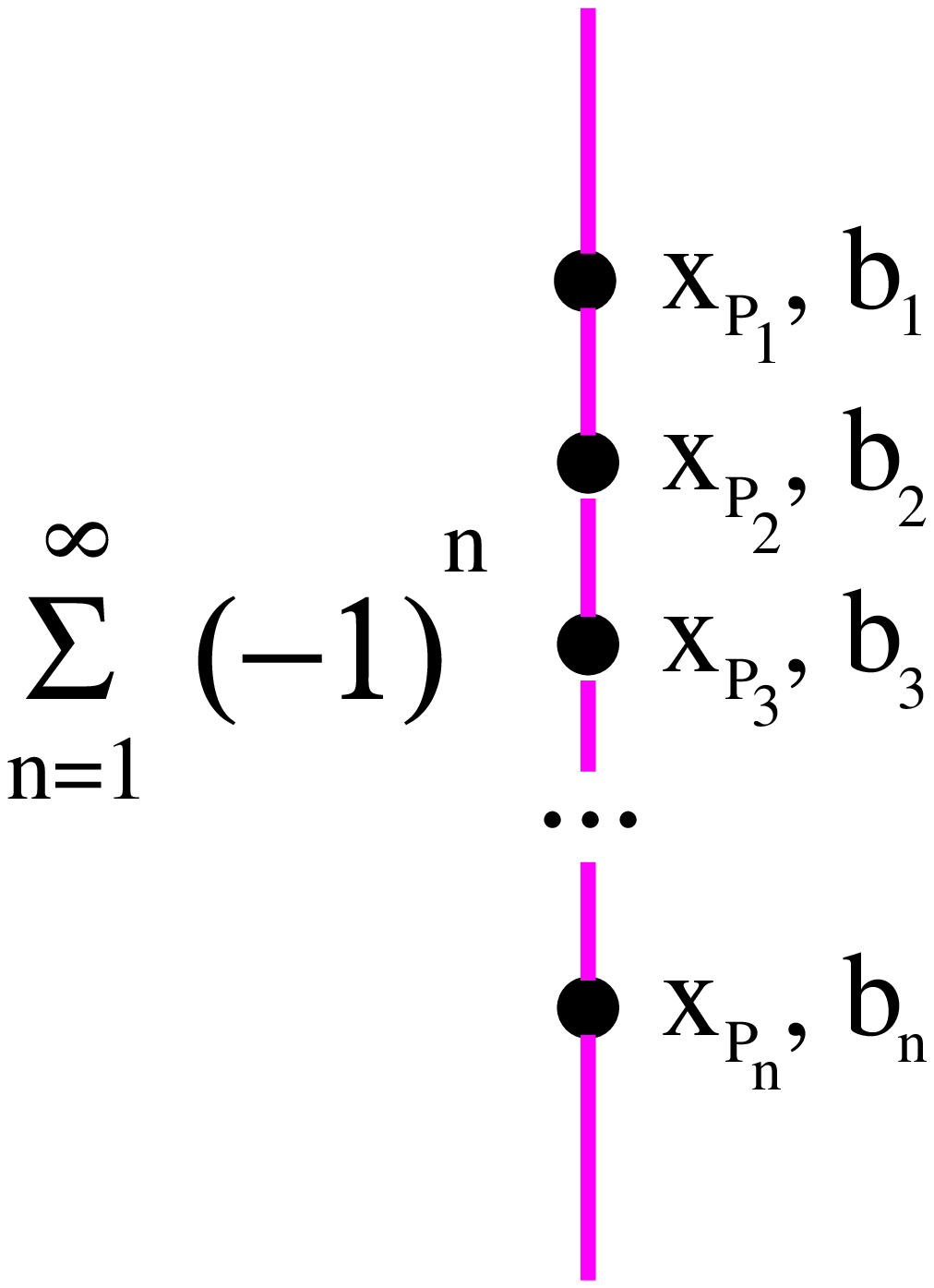}
  \end{center}
\caption{Left: lowest order enhanced graph. Right: dominant contributions
in the "dense" regime.\label{1-3P}}
\end{figure}

At sufficiently small energies one can restrict himself with just lowest in 
$r_{3{\rm P}}$ contribution (one multi-Pomeron vertex) -- Fig.~\ref{1-3P}(left),
which involves any transitions of $m \geq 1$ into $n \geq 1$ Pomerons minus
the Pomeron self-coupling ($m=n=1$). Going to higher energies it is the 
subtracted self-coupling graph which gives the largest contribution.
Adding also higher order diagrams, which can be obtained from the one in
Fig.~\ref{1-3P}(left) iterating the multi-Pomeron vertex in $t$-channel,
one obtains the "dense" limit (large $s$, small $b$) result as a sum over
corresponding self-coupling graphs -- Fig.~\ref{1-3P}(right).~\cite{kai86}
The latter corresponds to the usual (quasi-)eikonal Pomeron scheme, described
in the Introduction, however, with a re-normalized Pomeron intercept:
\begin{equation}
\tilde \alpha_{{\rm P}}(0)=
\alpha_{{\rm P}}(0)-4\pi\,r_{3{\rm P}}/\gamma_{\pi}
\end{equation}

In the general case one has to account for all essential enhanced diagrams
in order to obtain a smooth transition between the mentioned ``dilute'' 
(small $s$, large $b$) and ``dense'' limits. To this end we obtain the
so-called "fan" contribution, defined via a recursive equation of  
Fig.~\ref{ffan}(left),
and introduce also a ``generalized fan'' -- using a similar equation of
Fig.~\ref{ffan}(right); the difference between the two being due to vertexes 
with both ``fans'' connected to the projectile
and ones connected to the target:
\begin{figure}[ht]
\begin{center}
\includegraphics[
  width=6cm,
  height=2.5cm]{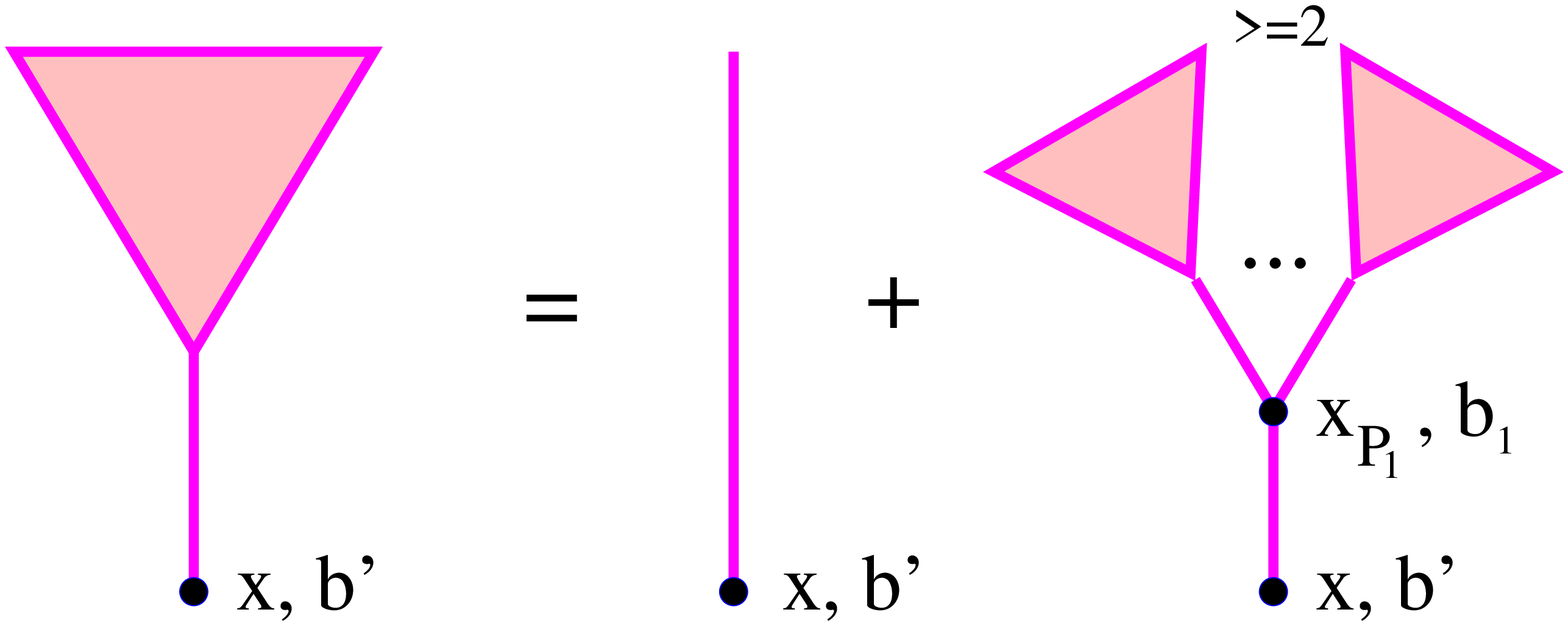}
  \hspace{2cm}
\includegraphics[%
  width=7cm,
  height=2.5cm]{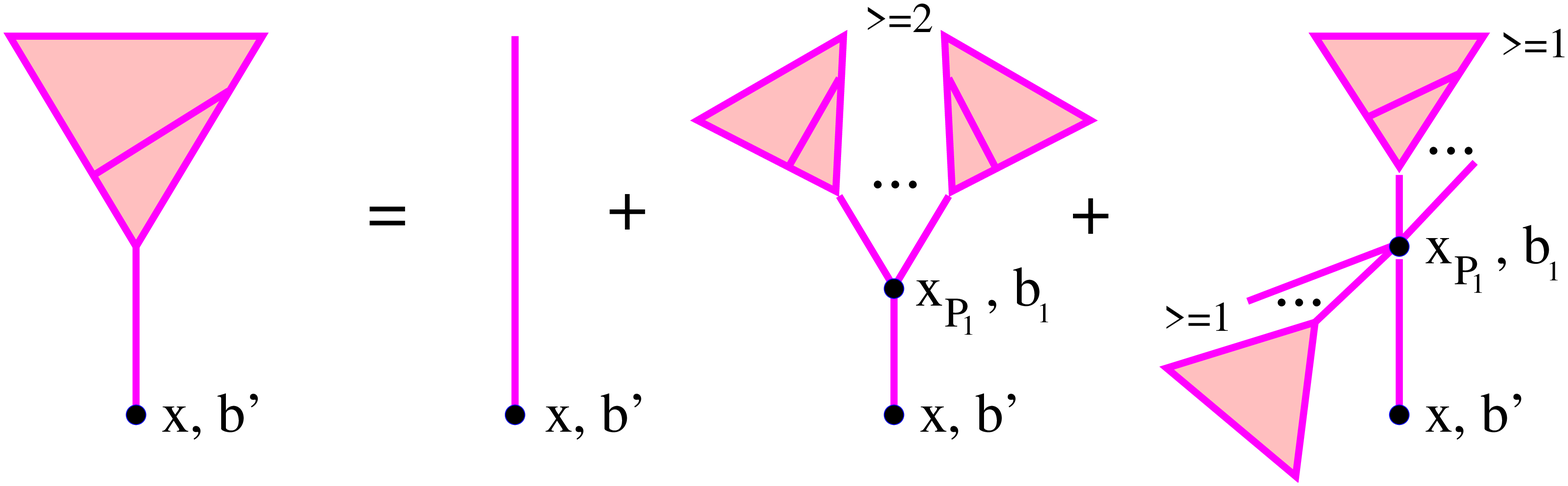}
    \end{center}
\caption{Recursive equations for the ``fan'' (left) and ``generalized fan'' (right)
 contributions.\label{ffan}}
\end{figure}

Then, neglecting so-called ``loop'' and ``chess-board'' graphs, whose contributions
always remain sub-dominant, and taking into consideration all other enhanced diagrams
of "net" type, with arbitrary topologies, one can express the full hadron-hadron
scattering amplitude via these elementary "building blocks" in a rather simple 
form:~\cite{erice} as the usual single Pomeron exchange, plus a sum over 
graphs with any number ($\geq 1$) of ``generalized fans'' connected 
to the projectile hadron and any number ($\geq 1$) of those connected to the target,
all coupled together in a single "central" vertex, minus corrections for Pomeron 
self-couplings and for double counting contributions. The obtained result indeed
interpolates smoothly between the two limited cases mentioned above: in the "dilute"
limit the multi-Pomeron part reduces to the lowest order contribution of 
Fig.~\ref{1-3P}(left), whereas in the "dense" regime one arrives to the asymptotic 
re-summation of Fig.~\ref{1-3P}(right). The same answer is obtained also for 
hadron-nucleus and nucleus-nucleus interactions, however, with an important difference
that both different  ("generalized") "fans" and different Pomerons in the same "fan"
contribution may couple to different projectile (target) nucleons. In case of nuclear collisions
the latter circumstance leads to an $A$-enhancement of corresponding screening corrections
in the "dilute" regime.

To describe particle production one has to consider all relevant unitarity cuts 
of this amplitude. Proceeding in the usual way, i.e.~applying AGK cutting 
rules~\cite{agk} and summing together contributions of cuts of certain
topologies, one can obtain positively defined probabilities for various configurations 
of the interaction, which allows to develop corresponding MC generation 
procedure and to treat non-linear interaction
effects explicitely in individual hadronic and nuclear collisions.~\cite{erice}

\section{Matching with QCD?}

  \begin{figure}[ht]
\begin{center}
\includegraphics[
  width=6cm,
  height=2.5cm]{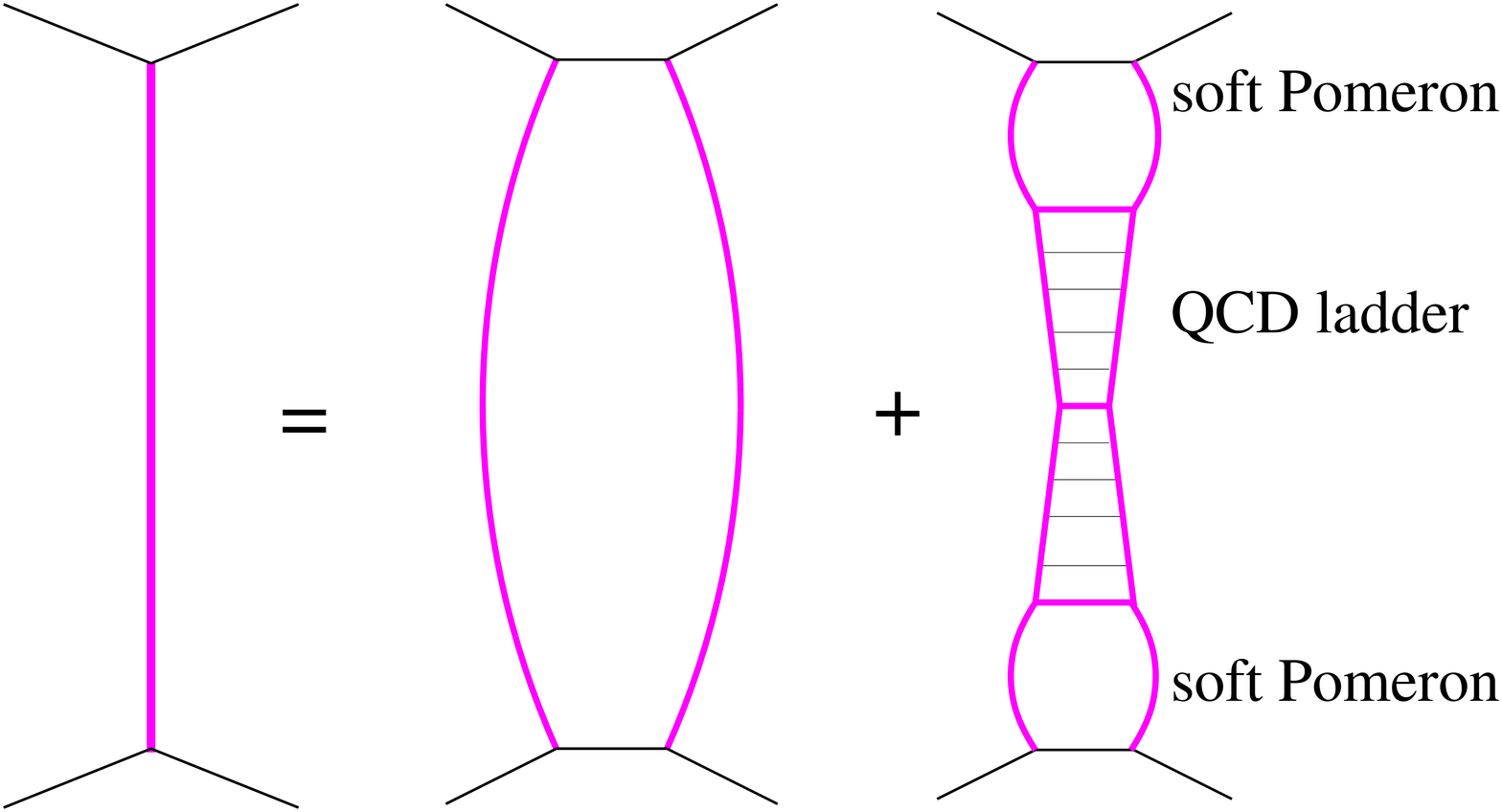}
  \hspace{1.cm}
\includegraphics[
  width=7cm,
  height=2.5cm]{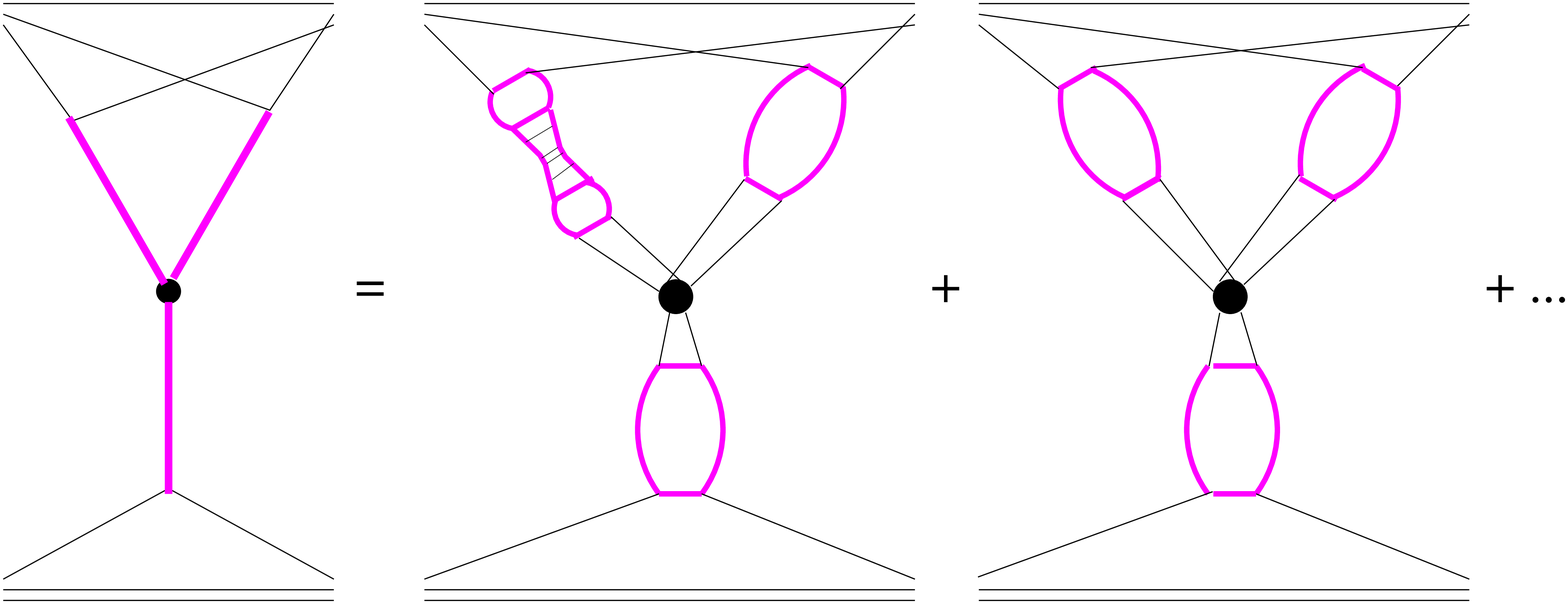}
    \end{center}
\caption{Left: a ``general Pomeron'' (l.h.s.) consists of the ``soft'' and
the ``semi-hard'' Pomerons -- correspondingly the 1st and the 2nd
contributions on the r.h.s. Right: contributions to the triple-Pomeron vertex
 from interactions between "soft" and "semi-hard" Pomerons.\label{genpom} }
\end{figure}
An essential drawback of the described scheme is that the high energy behavior of
hadron-hadron scattering amplitude is governed by a phenomenological Pomeron intercept,
not being constraint by perturbative QCD results. On the other hand, it lacks
a microscopic description of high $p_t$ parton processes and thus does not allow to 
calculate corresponding observables.

A possible solution could be to employ the phenomenological Pomeron description only
for "soft" cascades of partons of small virtualities $|q^{2}|<Q_{0}^{2}$,
while treating perturbative parton evolution at  $|q^{2}|>Q_{0}^{2}$ within pQCD
framework,  $Q_{0}^{2}$ being some cutoff for pQCD being applicable.
Without Pomeron-Pomeron interactions one thus
obtains the usual linear scheme  described in the Introduction,
however, based on the ``general Pomeron''. 
The latter consists of two contributions:~\cite{kal94,dre99}
phenomenological ``soft'' Pomeron for a pure non-perturbative process 
(all $|q^{2}|<Q_{0}^{2}$) and a so-called ``semi-hard Pomeron'', being a $t$-channel
iteration of the ``soft'' Pomeron and the QCD ladder, \footnote{A similar approach is the
"heterotic Pomeron" scheme.~\cite{tan94}} for a cascade which at least 
partly develops in the high virtuality region   
(some $|q^{2}|>Q_{0}^{2}$)  -- Fig.~\ref{genpom}(left).

 To account for corresponding non-linear effects we assume
that Pomeron-Pomeron interactions are dominated by partonic
processes at comparatively low virtualities,
 $|q^{2}|<Q_{0}^{2}$. Then we obtain precisely the scheme described in the
 preceeding Section, now based on the ``general Pomeron'' and with
multi-Pomeron vertexes involving only interactions
 between "soft" Pomerons or between ``soft ends'' of ``semi-hard Pomerons`` --
Fig.~\ref{genpom}(right). The parameters of such a scheme can be fixed based on data
on hadron-hadron cross sections, hadronic multi-particle production, and on the
measured total and diffractive proton structure functions.~\cite{erice}

\end{document}